\documentclass[%
 reprint,
 amsmath,amssymb,
 aps,
pra,
]{revtex4-1}

\usepackage{graphicx}
\usepackage{dcolumn}
\usepackage{bm}
\usepackage{braket}

\begin{document}

\preprint{APS/123-QED}

\title{Quantum correlations from dynamically modulated optical nonlinear interactions}

\author{Usman A. Javid$^{1,2}$}

\author{Qiang Lin$^{1,2,3}$}%
\email{qiang.lin@rochester.edu}
\affiliation{$^{1}$Institute Of Optics, University of Rochester, Rochester NY 14627, USA}
\affiliation{$^{2}$Center for Coherence and Quantum Optics, University of Rochester, Rochester NY 14627, USA}
\affiliation{$^{3}$Department of Electrical and Computer Engineering, University of Rochester, Rochester NY 14627, USA}

\date{\today}

\begin{abstract}
We investigate optical nonlinear interactions in a dynamic environment by studying generation of photons in spontaneous parametric down conversion inside a nonlinear cavity where the optical path length is periodically modulated in time. We show that the temporal dynamics of the cavity modify the nonlinear interaction and the generated continuous variable time-frequency entangled bi-photon state evolves into a tunable discrete higher dimensional state in the non-adiabatic modulation regime where the modulation time scales are much faster than the photon lifetime. In this regime, the system mimics effects of a quantum random walk in a photonic lattice with many associated effects including localized and delocalized wavefunctions of the generated photons. We also propose generation of time-frequency hyper-entangled states in the adiabatic limit. Our analysis shows that the proposed system is promising for applications in quantum simulation and information processing in the time-frequency domain.
\end{abstract}

\maketitle


\section{\label{sec:level1}Introduction}
Optical interactions in a cavity with time-varying boundary conditions have led to interesting results from both a fundamental and practical point of view \cite{review1}.  In cavity QED, optomechanical interaction \cite{optomechanics} between a quantized cavity field and a movable mirror has shown modifications of optical spectrum and coherence properties \cite{optospectrum1,optospectrum2,optospectrum3} as well as generation and control of exotic quantum states of light and the mechanical motion \cite{optomech1,optomech2,mechsqueeze}. In systems with single emitters coupled to modulated cavities, modification to spontaneous emission rates \cite{SPcontrol}, emission spectra and emitter-cavity coupling dynamics has been studied and tested \cite{atomcavity,emitterspectrum}. Cavities modulated by periodic signals have also been shown to be useful in optical isolation, non-reciprocity and topological photonics. \cite{isolation,nonreciprocal,topology}.

\par In this article, we investigate behavior of entangled bi-photon states generated inside a cavity containing a nonlinear medium where the optical path length is modulated by an arbitrary periodic signal. Studies have been done to some extent in modulated cavities with nonlinear media \cite{agatwal1,nonlinearmechanical}, however these systems rely on a resonant optomechanical interaction often requiring strong coupling at the single photon level and ultra-low temperatures. In order to arbitrarily modulate a cavity, non-resonant processes such as electro-optic effect, cross-phase modulation or free carrier dispersion are needed. Our treatment is also applicable to optomechanical systems when driven at the mechanical resonance frequency at room temperature where the quantum effects of the mechanical motion can be ignored. 

\par Optical parametric processes such as spontaneous parametric down conversion (SPDC) and spontaneous four-wave mixing (SFWM) are routinely used to generate entangled states of light. In the weak pumping regime, these processes produce pairs of photons with continuous variable (cv) entanglement in time and frequency \cite{entangled}. We show that this initial state evolves into a discrete higher dimensional state in frequency with tunable amplitudes in the resolved-sideband regime, where the modulation frequency exceeds the cavity linewidth. This results in the generated photons experiencing a time-varying cavity length. The dynamic system mimics features of quantum random walks \cite{random} including localized and delocalized wavefunctions. Such systems have been shown to be useful in quantum simulation \cite{sim1,sim2} and implementing information processing algorithms \cite{algo1,algo2}. However, unlike previous implementations of a quantum random walk where a quantum state is inserted into a lattice, here the photon emission process itself is randomized. We also propose a method to generate time-frequency bin hyper-entangled states in the unresolved-sideband (adiabatic) regime where the modulation frequency is much smaller than the cavity linewidth. These states can be turned into higher dimensional cluster states \cite{highD} for measurement based quantum computation \cite{oneway}. The proposed system can be implemented on a chip with a monolithic resonator carved from a nonlinear material.

 \begin{figure}
  \includegraphics[scale=1]{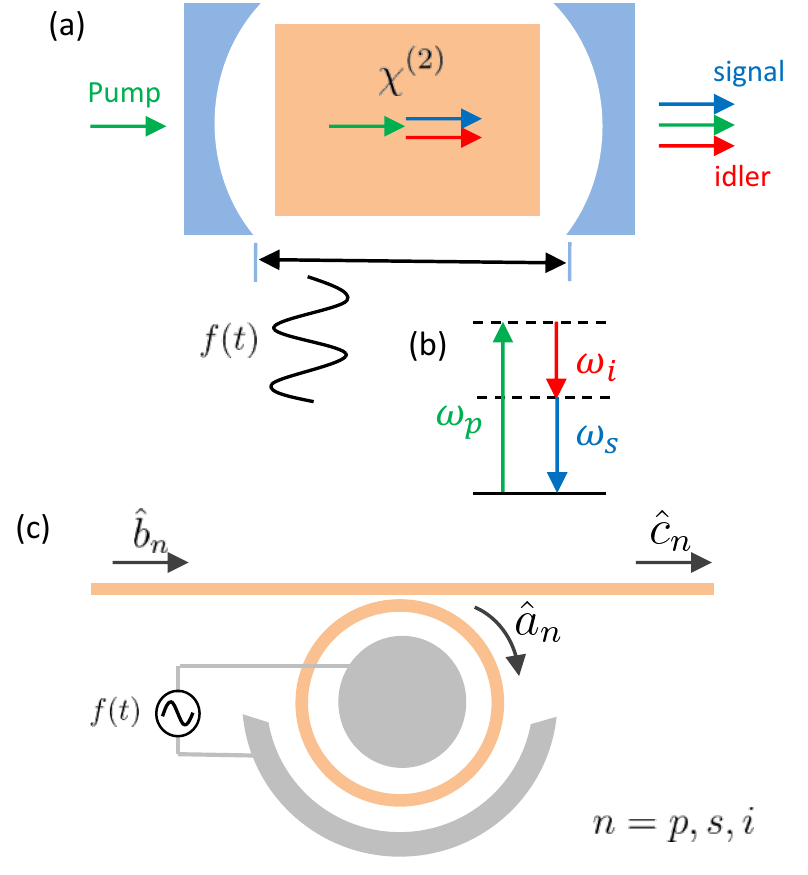}
  \centering
  \caption{Illustration of the proposed system. (a) An optical cavity with adjustable optical path length represented by $f(t)$ contains a nonlinear crystal with $\chi^{(2)}$ nonlinearity. A pump laser (green) drives an SPDC interaction inside the cavity to generate a pair of photons with frequencies that conserve energy as shown in the energy diagram in (b). (c) A practical scheme to implement the system in (a) using a ring resonator coupled to a waveguide with electrodes to change the refractive index using electro-optic effect.}
  \label{fig1}
\end{figure}

\section{\label{sec:level2}Theory}
Figure \ref{fig1}(a) illustrates the proposed system. A nonlinear medium is placed inside a cavity where the optical path length can be varied in time either by moving the cavity mirrors or by changing the refractive index of the medium. For simplicity, we take the case of SPDC but our treatment can be extended easily to any parametric nonlinear interaction. The cavity has three resonant modes phase and frequency matched for this interaction labeled as pump, signal and idler as shown in Fig. \ref{fig1}(b). A pump beam couples from one side of the cavity spontaneously generating a pair of photons with different frequencies. These photons then leak out of the cavity and are subsequently measured. Figure 1(c) shows a practical scheme to implement such system on a nanophotonic chip. A waveguide couples light into and out of a resonator etched from a medium with $\chi^{(2)}$ nonlinearity. Electro-optic effect is used to modulate the refractive index of the resonator by applying microwave signals to metallic electrodes in close proximity to the resonator. The Hamiltonian of the closed system can be described as
\begin{align}
\label{eqH}
    \hat{H}=\hat{H_0}+\hat{H}_{MOD}+\hat{H}_{NL}.
 \end{align}
In Eq. \eqref{eqH}, $H_0$ is the free field Hamiltonian given by
 \begin{align}
 \label{eqH0}
    \hat{H}_0=\sum_{n=p,s,i} \hbar \omega_{0n}\hat{a}^{\dag}_n \hat{a}_n,
 \end{align}
 where $a_n$ (n = p (pump), s (signal) and i (idler)) is the field operator inside the cavity as shown in Fig. 1(c) with unperturbed resonance frequency $\omega_{0n}$. $H_{MOD}$ is the part of the Hamiltonian that generates modulation of the cavity resonances. 
\begin{align}
\label{eqmod}
    \hat{H}_{MOD}=-\sum_{n=p,s,i}\hbar f(t) \hat{a}^{\dag}_n \hat{a}_n,\\
    \label{eqmod2}
    f(t)=\sum^{N}_{k=1}G_k \cos(k\Omega t+\phi_k).
 \end{align}
We have defined the modulation signal $f(t)$ in terms of its Fourier components with amplitudes $G_n$, phases $\phi_n$ and a fundamental frequency $\Omega$. We note that the Hamiltonian in Eq. \eqref{eqmod} is valid under the assumption that $G_n,n\Omega<<\omega_{0m}$ for all combinations of $m$ and $n$ and $n\Omega$ is far shorter than the cavity free spectral range (FSR) \cite{hamiltonian}. $H_{NL}$ is the nonlinear part of the Hamiltonian given by
\begin{align}
    \hat{H}_{NL}=\hbar g \hat{a}_p\hat{a}^{\dag}_s\hat{a}^{\dag}_i+h.c.,
 \end{align}
where $g$ is the nonlinear coupling strength. The system defined by the Hamiltonian $H$ is coupled to two reservoirs: the loss due to imperfections in the cavity walls constitutes the intrinsic cavity linewidth $\gamma_{0n}$ and the coupling waveguide introduces an additional external coupling loss $\gamma_{en}$. The Heisenberg-Langevin equations of motion for this open system are then expressed as
\begin{align}
    \frac{d\hat{a}_n}{dt}=\frac{1}{i\hbar}[\hat{a}_n,\hat{H}]-\frac{(\gamma_{0n}+\gamma_{en})}{2}\hat{a}_n+\sqrt{\gamma_{0n}}\hat{u}_n+i\sqrt{\gamma_{en}}\hat{b}_n,
 \end{align}
where $\hat{u}_n$ and $\hat{b}_n$ are the Langevin noise operators associated with the decay rates $\gamma_{0n}$ and $\gamma_{en}$ respectively. We will treat the reservoirs as broadband Markovian systems satisfying the commutation relations
\begin{align}
    [\hat{u}_m(t),\hat{u}^{\dag}_n(t')]=\delta_{mn}\delta(t-t'),\\
    [\hat{b}_m(t),\hat{b}^{\dag}_n(t')]=\delta_{mn}\delta(t-t').
 \end{align}
The pump and the modulation signal are strong compared to the generated signal and idler modes due to weak nonlinear coupling ($ga_p\ll\gamma_{0n}+\gamma_{en}$) and are treated classically. This also justifies ignoring effects of self- and cross-phase modulation between the three modes. The equations of motion for the open system then become
\begin{align}
\label{eqs1}
    \frac{da_p}{dt}&=-[i(\omega_{0p}-f(t))+\frac{\gamma_{tp}}{2}]a_p+i\sqrt{\gamma_{ep}}b_p(t) e^{-i\omega_p t},\\
    \label{eqs2}
    \frac{d\hat{a}_s}{dt}&=-[i(\omega_{0s}-f(t))+\frac{\gamma_{ts}}{2}]\hat{a}_s \nonumber\\ 
    &-iga_p\hat{a}^\dag_i+\sqrt{\gamma_{0s}} \hat{u}_s+i\sqrt{\gamma_{es}}\hat{b}_s, \\
    \label{eqs3}
    \frac{d\hat{a}_i}{dt}&=-[i(\omega_{0i}-f(t))+\frac{\gamma_{ti}}{2}]\hat{a}_i \nonumber \\
    &-iga_p\hat{a}^\dag_s+\sqrt{\gamma_{0i}} \hat{u}_i+i\sqrt{\gamma_{ei}}\hat{b}_i,
\end{align}
where $\gamma_{tn}=\gamma_{0n}+\gamma_{en}$ is the linewidth of the loaded cavity resonance and $\omega_p$ is the center frequency of the pump laser with temporal profile $b_p(t)$. Setting $\omega_{p}$ = $\omega_{0p}$ when the laser is tuned to the center of the pump resonance, we can make a sequence of transformations to simplify these equations
\begin{align}
\label{eqtransI}
    \hat{O}_n=\hat{O}^{'}_n \times exp[-i(\omega_{0n}t], \quad (\hat{O}=\hat{a},\hat{b},\hat{u}).
 \end{align}
 This transformation puts the system into a rotating frame with a frequency matching the corresponding resonance frequency for each mode. We will now make a second transformation
 \begin{align}
\label{eqtrans}
    \hat{a}_n=\hat{a}^{'}_n \times exp[ih(t)],
 \end{align}
where
\begin{align}
    h(t)=\int^{t}_{-\infty} f(t^{'}) dt^{'}.
 \end{align}
Plugging these transformations in Eqs. \eqref{eqs1} to \eqref{eqs3} and assuming the unperturbed resonance frequencies are perfectly matched for SPDC ($\omega_{0p}$ = $\omega_{0s}$ + $\omega_{0i}$), we get
\begin{align}
\label{sys1}
    \frac{da^{'}_p}{dt}&=-\frac{\gamma_{tp}}{2}a^{'}_p+i\sqrt{\gamma_{ep}}b_p(t)e^{-ih(t)},\\
    \label{sys2}
    \frac{d\hat{a}^{'}_s}{dt}&=-\frac{\gamma_{ts}}{2}\hat{a}^{'}_s + (-iga^{'}_p\hat{a}_{i}^{'\dag}
    +\sqrt{\gamma_{0s}} \hat{u}_s+i\sqrt{\gamma_{es}}\hat{b}_s)e^{-ih(t)}, \\
    \label{sys3}
    \frac{d\hat{a}^{'}_i}{dt}&=-\frac{\gamma_{ti}}{2}\hat{a}^{'}_i 
    +(-iga^{'}_p\hat{a}_{s}^{' \dag}+\sqrt{\gamma_{0i}} \hat{u}_i+i\sqrt{\gamma_{ei}}\hat{b}_i)e^{-ih(t)},
\end{align}
 where we have kept the prime notation (') for the transformation in Eq. \eqref{eqtrans} only for simplicity. By making the transformation defined by Eq. \eqref{eqtrans}, we have shifted the system into a rotating phase modulated frame. In such a frame, a phase modulated wave would appear to be static. The transformed set of equations resemble a system with static resonances driven by a phase modulated input and the actual solution for the field in the modulated cavity differs from the transformed system only by a factor of a harmonic phase.  
We can now solve the system of Eqs. \eqref{sys1} to \eqref{sys3}.

\subsection{Steady state solution for pump} \label{secpump}
Eq. \eqref{sys1} can be exactly solved in the frequency domain by expanding $exp[-ih(t)]$ as a sum of harmonics. 
\begin{align}
\label{eqh1}
   exp[-ih(t)]&=exp[-i \sum^{N}_{k=1} \frac{G_k}{k\Omega} \sin(k\Omega t+\phi_k)]\nonumber\\&=\sum^{\infty}_{l=-\infty} f_l e^{-il\Omega t},
\end{align}
\begin{align}
   \label{eqh2}
   f_l=&\sum^{\infty}_{m_2,m_3,...,m_N=-\infty} J_{m_1}(\frac{G_1}{\Omega}) J_{m_2}(\frac{G_2}{2\Omega})...\times J_{m_N}(\frac{G_N}{N\Omega}) \nonumber\\ 
   &\times exp[-i\sum^{N}_{q=1} m_q \phi_q], \quad m_1=l-\sum^{N}_{r=2}r m_{r} ,
\end{align}
where the integer $N$ determines the number of sinusoidal signals used in the periodic signal from Eq. \eqref{eqmod2} and $J_m(x)$ is the Bessel function of the first kind and order $m$. The steady state solution for the pump can be obtained by taking Fourier transform of Eq. \eqref{sys1} and using Eq. \eqref{eqh1}.
\begin{align}
\label{eqpump}
    a^{'}_p(\omega)=\frac{i\sqrt{\gamma_{ep}}\sum^{\infty}_{l=-\infty}f_l b_p(\omega-l\Omega)}{\gamma_{tp}/2-i\omega},
\end{align}
where we have defined the Fourier transform as
\begin{align}
    a(\omega)=\int^{\infty}_{-\infty} a(t)e^{i\omega t} dt.
\end{align}
The solution for the inter-cavity pump field is a sum of shifted copies of the pump spectrum scaled by a lorentzian. This is not surprising since in this frame, the input pump field is phase modulated. The complete solution for the pump as well as the signal and idler modes after transforming back from the phase modulated frame will be
\begin{align}
\label{eqspec}
    a_n(\omega)=\sum^{\infty}_{l=-\infty} f^{*}_l a^{'}_n(\omega+l\Omega),
\end{align}
which is simply a Fourier transform of Eq. \eqref{eqtrans}
\begin{figure*}
\includegraphics[scale=1.1]{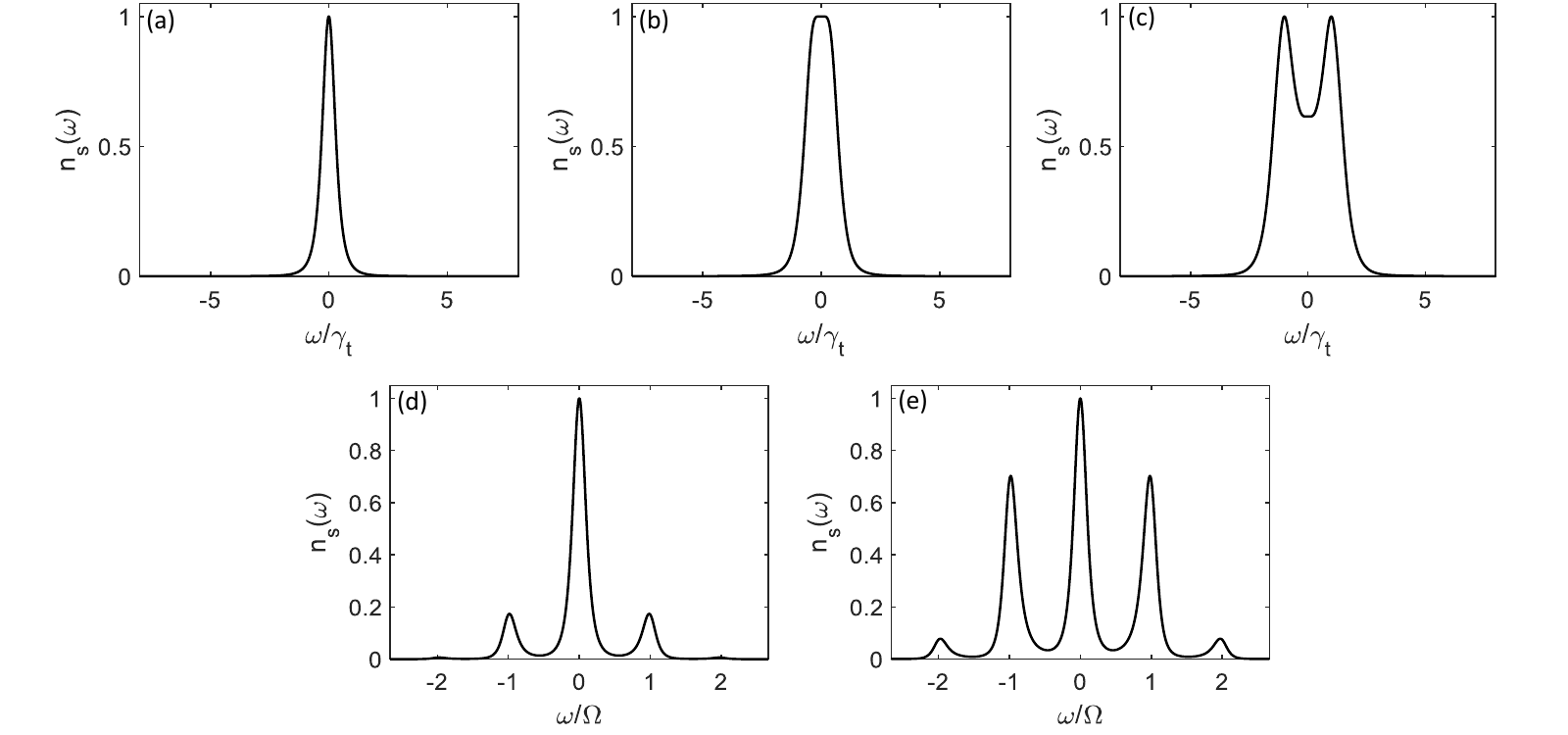}
\centering
\caption{\label{fig:wide}Normalized spectrum of the signal photons with modulation parameters $\Omega/\gamma_t$ and $G/\gamma_t$ equal to: (a) 0.5 and 0. (b) 0.5 and 1.25. (c) 0.5 and 3. (d) 3 and 0.75. (e) 3 and 1.25. The emergence of peaks in (c) is due to the sinusoidal form of the modulation as the time derivative is zero at the extreme end of the oscillation.}
\label{fig2}
\end{figure*}

\subsection{Perturbative treatment for signal and idler modes} \label{secpert}
In the weak pumping regime, we can treat the nonlinear part of the system as a small perturbation. We will solve the system perturbatively in the frequency domain \cite{perturb} to the first order where the nonlinear interaction generates at most two photons. To zeroth order, the idler equation Eq. \eqref{sys3} can be expressed as:
\begin{align}
\label{eq0}
    \frac{d\hat{a}^{'0}_i}{dt}&=-\frac{\gamma_{ti}}{2}\hat{a}^{'0}_i
    +(\sqrt{\gamma_{0i}} \hat{u}_i+i\sqrt{\gamma_{ei}}\hat{b}_i)e^{-ih(t)}.
\end{align}
Then the signal equation to first order is obtained by plugging the solution of Eq. \eqref{eq0} into Eq. \eqref{sys2}.
\begin{align}
\label{eq1}
    \frac{d\hat{a}^{'1}_s}{dt}&=-\frac{\gamma_{ts}}{2}\hat{a}^{'1}_s + (-iga^{'}_p\hat{a}_{i}^{' 0 \dag}
    +\sqrt{\gamma_{0s}} \hat{u}_s+i\sqrt{\gamma_{es}}\hat{b}_s)e^{-ih(t)}.
\end{align}
A similar procedure can be done for the first order solution of the idler and is equivalent to swapping the subscript $s$ with $i$ in the solution. We will solve Eqs. \eqref{eq0} and \eqref{eq1} in the frequency domain to obtain their steady state solution. To simplify things, we will assume $\gamma_{0p}$ = $\gamma_{0s}$ = $\gamma_{0i}$ = $\gamma_0$, $\gamma_{ep}$ = $\gamma_{es}$ = $\gamma_{ei}$ = $\gamma_e$ and $\gamma_{0}$ + $\gamma_{e}$ = $\gamma_{t}$. The resulting solutions, ignoring the 0, 1 notation for simplicity, and transforming back from the phase modulation frame using Eq. \eqref{eqspec}, are calculated to be
\begin{widetext}
\begin{align}
\label{eqsol}
    \hat{a}_{s,i}(\omega)=\sum^{\infty}_{j,k,l=-\infty} A_{j,k,l}(\omega)\int^{\infty}_{-\infty} d\omega^{'} \frac{\hat{\xi}^{\dag}_{i,s}(-\omega^{'}-l\Omega)a^{'}_p(\omega-\omega^{'}-(j-k)\Omega)}{\gamma_t/2-i\omega^{'}}+\sum^{\infty}_{j,k=-\infty} B_{j,k}(\omega)\hat{\xi}_{s,i}(\omega-(j-k)\Omega),\\
    \text{where} \quad  A_{j,k,l}(\omega)=-\frac{igf_{j}f^{*}_{k}f^{*}_{l}}{\gamma_t/2-i(\omega+k\Omega)}, \quad B_{j,k}(\omega)=\frac{f_{j}f^{*}_{k}}{\gamma_t/2-i(\omega+k\Omega)} \quad \text{and} \quad \hat{\xi}_{s,i}(\omega)=\sqrt{\gamma_0}\hat{u}_{s,i}(\omega)+i\sqrt{\gamma_e}\hat{b}_{s,i}(\omega). \nonumber
\end{align}
\end{widetext}
In this solution, $a_p^{'}(\omega)$ is the pump spectrum in the modulation frame given by Eq. \eqref{eqpump}. The operators $\hat{\xi}_n(\omega)$ ($n$ = $s$, $i$) are the sum of the noise operators of the two reservoirs and satisfy the commutation relation
\begin{align}
\label{eqcom}
   [\hat{\xi}_{m}(\omega),\hat{\xi}^{\dag}_{n}(\omega^{'})]=2\pi \gamma_t \delta_{mn}\delta(\omega-\omega^{'}).
\end{align}

\section{Results}
We can define the operators for the field transmitted out of the cavity in Fig. \ref{fig1}(c) using cavity input-output theory as
\begin{align}
\label{eqout}
    \hat{c}_{s,i}(\omega)=\hat{b}_{s,i}(\omega)+i\sqrt{\gamma_e}\hat{a}_{s,i}(\omega).
\end{align}
We now have all the necessary relationships to calculate the spectral profiles and coherence properties of the emitted photons.

\subsection{Single Channel Spectrum}
The spectrum of the signal mode is given by
\begin{align}
    n_s(\omega)=\langle\hat{c}^{\dag}_{s}(\omega)\hat{c}_{s}(\omega)\rangle.
\end{align}
We will evaluate $n_s(\omega)$ for a pump with a Gaussian spectrum given by $b_p(\omega)=b_{0}exp[-\omega^2/\sigma^2]$.  Figure \ref{fig2} plots the signal spectrum using the results from Eq. \eqref{eqsol} in the limit $\sigma<<\Omega,\gamma_t$ for a simple case where $N$ = 1 and $f(t)=G\cos{(\Omega t)}$. We can divide the results into two regimes of operation for the system: The unresolved-sideband or adiabatic regime where $\Omega<\gamma_t$ and resolved-sideband regime where $\Omega>\gamma_t$. Figure \ref{fig2}(b)-(c) plot the spectrum in the adiabatic regime with $\Omega/\gamma_t$ = 0.5. We see that as the modulation amplitude is increased from zero, the spectrum broadens. This happens due to the back and forth oscillations of the resonances since the driving frequency is small enough for the cavity to follow. Therefore the center frequencies of the generated photons oscillate in time and the total spectrum shows this as a broadening. This motion of the spectrum in time can be used for quantum frequency conversion \cite{Tang}. It is important to note that if the modulation frequency is far shorter than the photon lifetime, the broadening will disappear since the cavity moves too slow for the pump energy to survive long enough to get to the extreme end of the oscillation. Therefore all the photons are generated close to the center of the unperturbed resonance frequencies $\omega_{0n}$ ($n$ = $s$, $i$). This is equivalent to a pulsed pump driving the nonlinear interaction inside a static cavity. Figure \ref{fig2}(d)-(e) show the results in the resolved-sideband regime. Here we see creation of sidebands at integer multiples of the modulation frequency $\Omega$ indicating a three-wave sum/difference frequency interaction of the photons with the modulation signal. The strength of these sidebands can be controlled with the modulation amplitude.

\begin{figure}
\includegraphics[scale=0.85]{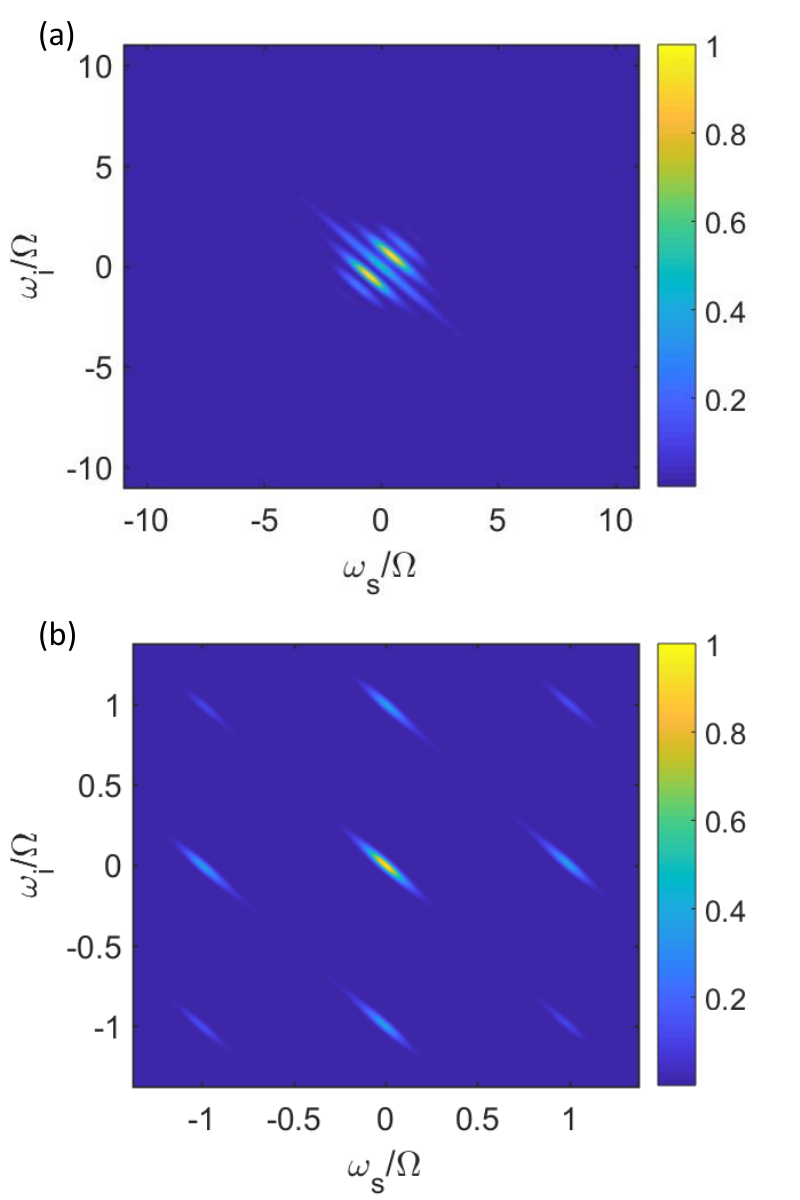}
\centering
\caption{Joint spectral intensity $|S(\omega_s,\omega_i)|^2$ normalized to $|S(0,0)|^2$ for $f(t)$ = $G\cos(\Omega t)$ with parameters $\Omega/\gamma_t$ and $G/\gamma_t$ equal to  (a) 0.5 and 0.5. (b) 4 and 0.5}
\label{fig3}
\end{figure}

\subsection{Joint Spectrum}
In the weak pumping regime, the quantum state of the light emitted from such a nonlinear interaction can be given as \cite{entangled}
\begin{align}
    |\psi\rangle=C_0|0\rangle + \int\int d\omega_s d\omega_i S(\omega_s,\omega_i)|\omega_s\rangle|\omega_i\rangle,
\end{align}
where $|0\rangle$ is the vacuum state and $S(\omega_s,\omega_i)$ is the complex joint spectral amplitude. The generated photons are in a continuous variable entangled state. If we evaluate the expectation value of the operator $\hat{c}_s(\omega_s)\hat{c}_i(\omega_i)$ in this state, we get
\begin{align}
    \langle \psi| \hat{c}_s(\omega_s)\hat{c}_i(\omega_i) | \psi\rangle= C_{0}S(\omega_s,\omega_i).
\end{align}
This gives the complex joint spectral amplitude correct up to a normalization constant. Therefore in the Heisenberg picture, the expectation should give the same result. We can evaluate $S(\omega_s,\omega_i)$ using Eqs. \eqref{eqsol}, \eqref{eqcom} and \eqref{eqout} to obtain

\begin{align}
\label{eqS}
     S(\omega_s,\omega_i) &\propto 2\pi\gamma_{e}\sum^{\infty}_{j,k,l=-\infty}\frac{A_{j,k,l}(\omega_i)a^{'}_p(\omega_s+\omega_i-(j-k-l)\Omega)}{\gamma_{t}/2+i(\omega_s+l\Omega)} \nonumber\\
    &-2\pi\gamma_{t}\gamma_{e}\sum^{\infty}_{j.k.l.m.n-\infty}B_{j,k}(\omega_s)A_{l,m,n}(\omega_i) \nonumber\\
    &\times\frac{a^{'}_p(\omega_s+\omega_i-(j-k+l-m-n)\Omega)}{\gamma_{t}/2+i(\omega_s-(j-k-n)\Omega)}.
\end{align}
Just like the spectrum for the signal mode, the joint spectrum also has lorentzians shifted by integer multiples of the modulation frequency $\Omega$. Figure \ref{fig3} plots the joint spectral intensity given by $|S(\omega_s,\omega_i)|^2$ for $\sigma/\gamma_t=0.1$ for both the unresolved- and resolved-sideband regimes. There are two distinct phenomena caused by the modulated cavity boundaries that lead to the joint spectrum of the form in Eq. \eqref{eqS}. Firstly, the pump becomes phase modulated and each of its sidebands can generate photon pairs. This leads to the narrowly spaced spectral lines in Fig. \ref{fig3}(a) in the adiabatic regime. This result is similar to generation of photon pairs in  a static cavity with a phase modulated pump. Secondly, the scattering of the generated photons into sidebands themselves contributes to the energy in these shifted spectral lines. Both of these phenomenon are at work and can interfere. For instance, a photon pair located at the first sideband $\omega_s/\Omega=1$, $\omega_i/\Omega=0$ can get here by being generated at $\omega_s/\Omega=0$, $\omega_i/\Omega=0$ and then the signal scatters into this sideband, or both the photons can be directly generated here from a pump sideband shifted by an amount $\Omega$ from the pump center frequency. These interference effects can allow changing the strength of the generated sidebands and is discussed in section \ref{sec:HDE}. This also sheds light on the appearance of photons at energies away from the diagonal line ($\omega_s +\omega_i=0$) on the signal-idler frequency plain,  a feature not available in photon pair generation in static environments.

\begin{figure*}
\includegraphics[scale=1]{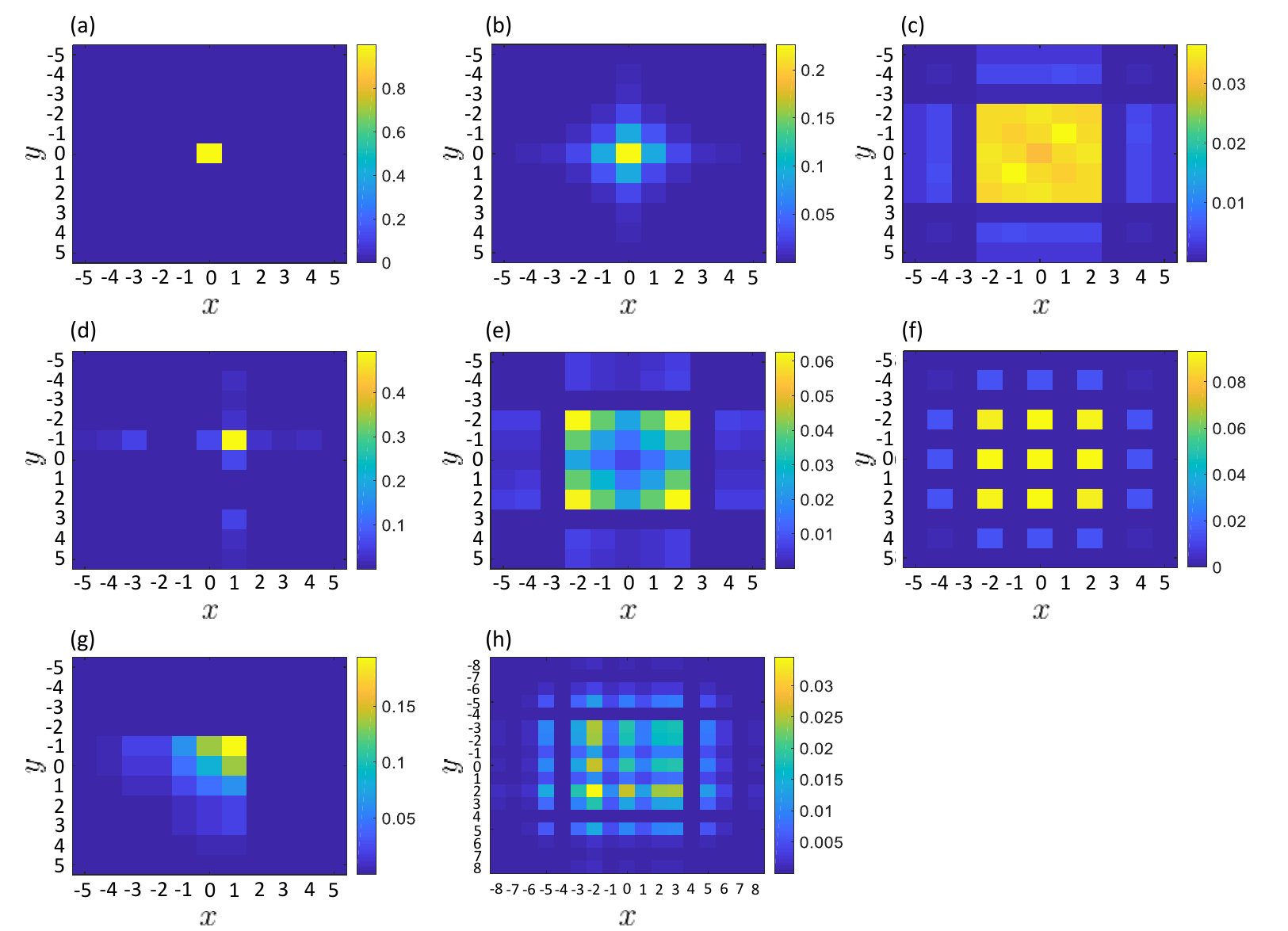}
\centering
\caption{Quantum state coefficients $|C_{x,y}|^2$ with $N$ = 3 for various values of the six parameters $G_1,G_2,G_3,\phi_1,\phi_2,$ and $\phi_3$ given in Table \ref{tab1}.}
\label{fig4}
\end{figure*}

\begin{table}
\caption{Modulation parameters for Fig. 4}
\label{tab1}
\begin{ruledtabular}
\begin{tabular}{c c c c c c c c c c}
\textrm{Figure 4}&
\textrm{$G_1/\Omega$}&
\textrm{$G_2/\Omega$}&
\textrm{$G_3/\Omega$}&
\textrm{$\phi_1/\pi$}&
\textrm{$\phi_2/\pi$}&
\textrm{$\phi_3/\pi$}&\\
\colrule
(a) & 0 & 0 & 0 & 0 & 0 & 0 \\
(b) & 1.1 & 0 & 1.3 & 1 & 0 & 0 \\
(c) & 1.658 & 0 & 2.04 & 0.703 & 0.921 & 1.137 \\
(d) & 2 & 1.811 & 1.757 & 0.287 & 0.671 & 1 \\
(e) & 1.952 & 0 & 2.233 & 0 & 0 & 1 \\
(f) & 0 & 2.85 & 0 & 0 & 0 & 0 \\
(g) & 1.36 & 0.85 & 0.85 & 0 & 0 & 0 \\
(h) & 0.632 & 2.5 & 3.5 & 0.267 & 0.17 & 1 \\
\end{tabular}
\end{ruledtabular}
\end{table}

\subsection{\label{sec:HDE}Higher Dimensionality and Random Walks}
\begin{figure}
\includegraphics[scale=1]{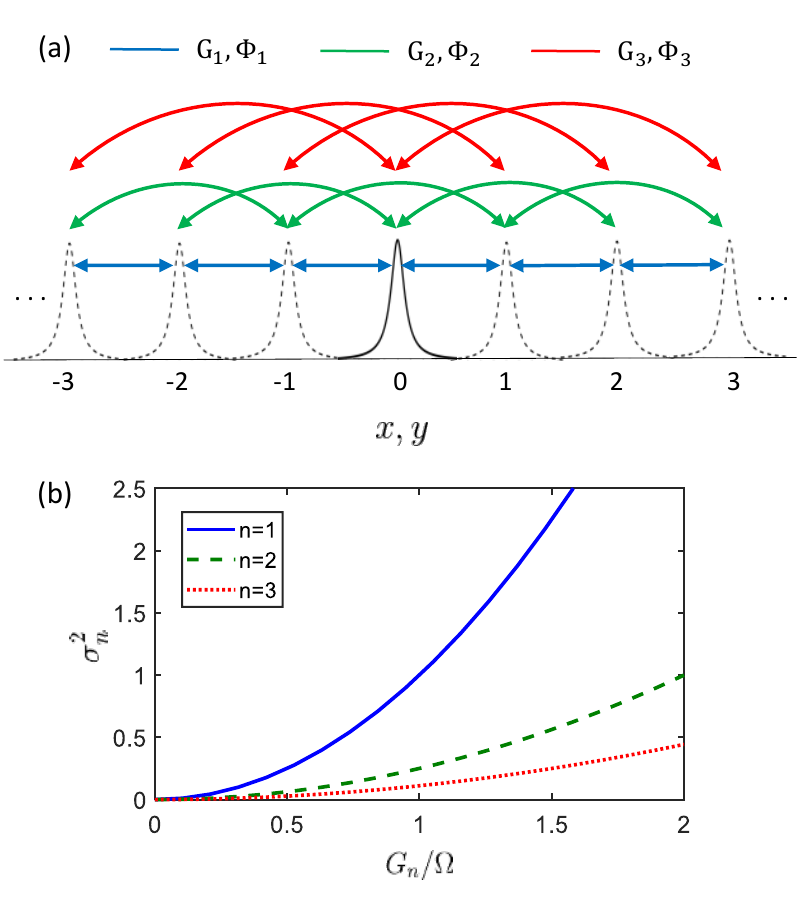}
\centering
\caption{(a) Illustration of interference pathways of the generated photons for the first three harmonics of the modulation signal. Each path has an amplitude $G_n$ and phase $\phi_n$ ($n$ = 1, 2, 3). (b) Variance of the bi-photon wavefunction for each harmonic as the function of the modulation amplitude while the other two amplitudes are set to zero.}
\label{fig5}
\end{figure}

The results shown in Fig. 3(b) motivate us to write the $S(\omega_s,\omega_i)$ as a coherent sum of joint spectra of static cavities shifted in frequency as
\begin{align}
|\psi\rangle=\sum_{x,y}C_{x,y}\int\int S_{0}(\omega_s+x\Omega,\omega_i+y\Omega)|\omega_s\rangle|\omega_i\rangle,
\end{align}
where we have ignored the vacuum contribution and $S_{0}(\omega_s+x\Omega,\omega_i+y\Omega)$ is the joint spectral amplitude of the system when the modulation signal is turned off centered at $(-x\Omega,-y\Omega)$. This is a discrete higher dimensional two-party state in frequency of the form
\begin{align}
\label{eqdim}
|\psi\rangle=&\sum_{x,y} C_{x,y}|x,y\rangle,\\
\text{where} \quad |x,y\rangle=\int\int S_{0}(\omega_s&+x\Omega,\omega_i+y\Omega)|\omega_s\rangle|\omega_i\rangle.
\end{align}
The coefficients of the state are given by $C_{x,y}=\langle x,y|\psi\rangle$. Figure \ref{fig4} plots the modulus squared of these coefficients numerically calculated for $N$ = 3 in the resolved-sideband regime. We have set $\Omega/\gamma_{t}$ = 10 to ensure that the basis states of $|\psi\rangle$ have no overlap. By setting $N$ = 3, there are 6 parameters ($G_1,G_2,G_3,\phi_1,\phi_2,\phi_3$) that can be adjusted making the state tunable. Therefore frequency domain control of the quantum state can be turned into a parameter optimization problem. As an example, we have optimized several states. For instance, Fig. \ref{fig4}(c) shows a near equal superposition for 25 basis states closest to the center. Figure \ref{fig4}(d) shows that at certain values of these parameters, the system resembles a frequency shifted version of an un-modulated bi-photon source. Therefore this system can be used as a tunable two-mode frequency shifter useful in spectroscopy with squeezed states \cite{squeeze}. The ability to tune these coefficients is due to inteferences between multiple pathways that can lead up to a particular frequency bin for the two photons. Figure \ref{fig5}(a) shows these pathways for $N$ = 3. We can imagine each frequency bin as a node on a two dimensional lattice formed by the frequencies of the two photons. The number of modulation signals applied to the resonator determines the connectivity between the nodes. For $N$ = 1, the lattice only has nearest neighbour connectivity shown as blue arrows in Fig. \ref{fig5}(a) and the higher harmonics of the modulation signal create connections with distant neighbours. This creates a large number of paths with adjustable amplitudes and phases, giving control over the resulting interference. During their lifetime inside the cavity, the generated photons can scatter between these nodes. Therefore this system closely resembles a quantum random walk of an entangled state in a photonic lattice \cite{walk1,walk2,walk3,walk4}. These interference effects can generate walks with large asymmetric spread (Fig. \ref{fig4}(h)) that mimic a disordered medium, directional spread (Fig. \ref{fig4}(g)), or highly localized wavefunctions (Fig. \ref{fig4}(d)) even with large amplitudes (see Table \ref{tab1} for corresponding modulation amplitudes and phases). In order to characterize the spread of the random walk for each harmonic, we can calculate the variance of the output wavefunction gives as
\begin{align}
\sigma_{n}^2=\frac{\sum_{x,y} d_{n}^{2}(x,y)|C_{x,y}|^2}{\sum_{x,y}|C_{x,y}|^2},
\end{align}
where $\quad d_{n}(x,y)=\sqrt{(x/n)^{2}+(y/n)^{2}}$ is the distance of each bin from the center normalized such that adjacent bins connected by the n-th harmonic of the modulation signal are separated by a distance of 1 in each axis and the state coefficients $C_{x,y}$ are calculated by setting all the other harmonics to zero. The results are plotted in Fig. \ref{fig5}(b) for the first three harmonics. The plots show quadratic scaling and the variance decrease for a fixed modulation amplitude as we move to higher harmonics. This is also evident from the solution of the optical field inside the cavity in Section \ref{secpump} where we show that the strength of the sidebands scale as the n-th order Bessel function of $G_n/n\Omega$ making the sidebands smaller as $n$ increases. The tunneling rates of the photons between adjacent bins for each harmonic are determined by the corresponding modulation amplitude $G_n$ (see Appendix \ref{AppA} for details) and provide control of the random walk, a feature of frequency domain random walks that sets them apart from other photonic implementations.

\par Another unique feature of this system  is that the interference pathways are not just scattering events of photons in an otherwise passive lattice, rather the photon creation process itself is distributed in the lattice due to the sidebands of the pump which can generate photons away from the central lattice site ($x,y$ = 0). Therefore the photonic lattice is an active grid of parametric photon pair sources, locally generating entangled photons with SPDC with controllable strengths and phases and scattering them to other sites with the modulation signal. Since the modulation to the bi-photon spectrum is symmetric, the state at the output is separable. However, it is important to note that this separability is only in the discrete frequency space and a continuous variable entanglement is still present within the frequency bins due to the generation process. It is interesting to think about these correlations within the bins as another set of dimensions to the lattice. In lattices constructed by cavity modes, the detuning of the modulation signal from the center of the optical resonance can be used to create more dimensions \cite{detune}. This system naturally gives a continuum of dimensions to explore complex lattice structures, a topic for future communications.

\subsection{Hyper-Entanglement with Chirped Photons}
\begin{figure}
\includegraphics[scale=0.8]{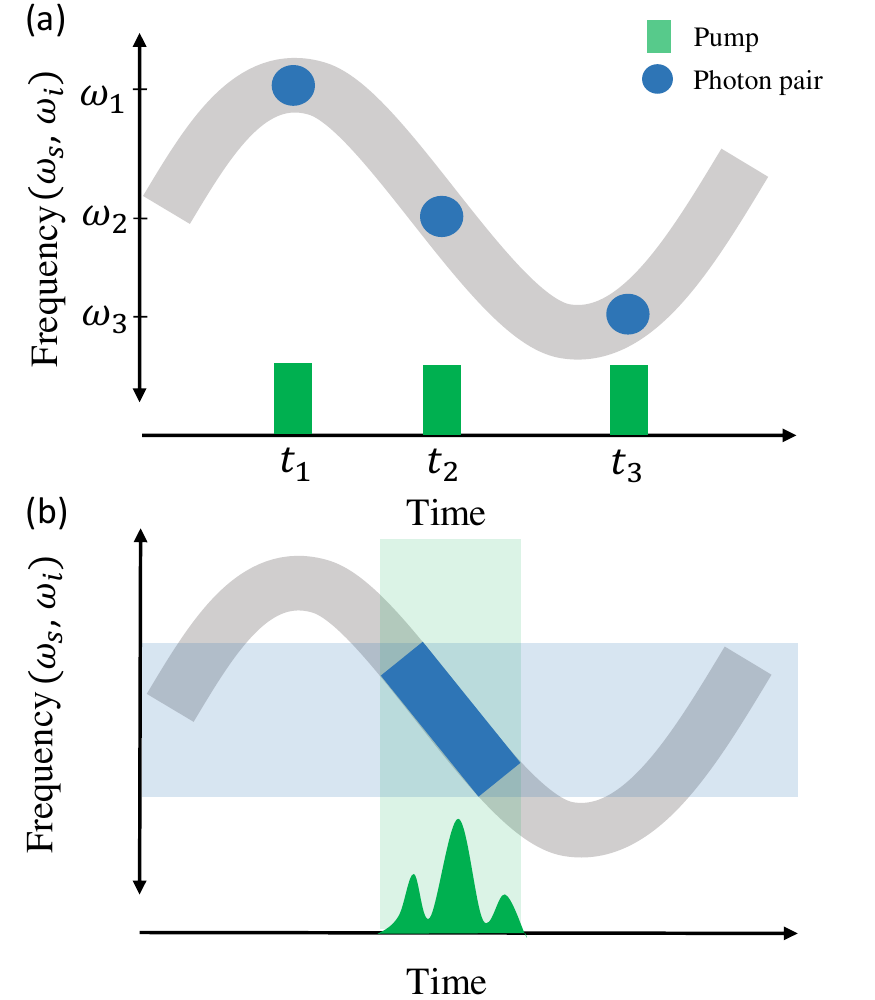}
\centering
\caption{Proposal for generation of time-frequency hyper-entanglement in (a) discrete and (b) continuous variable regime. The gray sinusoidal curves track the motion of the signal/idler resonances in time and blue regions on the curves point out the position of the photon pairs on the time-frequency plain generated by pump pulses shown in green.}
\label{fig6}
\end{figure}

Like the resolved sideband regime, the adiabatic regime can also generate complex quantum states of light. If we set $\Omega \ll \gamma_t$, then the cavity resonances slowly oscillate back and forth in time. However, since the photon lifetime is much smaller than the modulation period, the spectrum of the emitted photons are time tagged. We can then use a tunable pulsed laser source in a standard time-bin entanglement scheme \cite{timebin} with the laser locked to the pump center frequency. By choosing the arrival time and number of pump pulses, we can create states of the form
\begin{align}
|\Psi\rangle=&|\omega_1,t_1\rangle_{s}|\omega_1,t_1\rangle_{i}+|\omega_2,t_2\rangle_{s}|\omega_2,t_2\rangle_{i} \nonumber\\
&+|\omega_3,t_3\rangle_{s}|\omega_3,t_3\rangle_{i}.
\end{align}
Figure \ref{fig6}(a) shows the time and frequency bins associated with this state. These states are hyper-entangled in discrete time and frequency bins. States like these can be turned into a higher-dimensional cluster state \cite{highD} for measurement-based quantum computation \cite{oneway} where the two degrees of freedoms (time and frequency) are treated as independent parties to a cluster state. It is important to note that increasing the dimensionality will require large modulation amplitudes which have been achieved in many materials \cite{Tang,electrooptic} and a frequency modulated laser locked to the center of the moving pump resonance. Given a 1 ns average photon lifetime of typical microcavities, a delay on the order of tens of nanoseconds between pump pulses would be enough to generate this state without any overlap between time and frequency bins and loss of fidelity. We can also envision a continuous variable analog of this system with a pump pulse of an arbitrary shape as shown in Fig. \ref{fig6}(b). The corresponding entangled state will be of the form

\begin{align}
    |\Psi\rangle=P(t)\int\int d\omega_s d\omega_i S(\omega_s+\alpha t, \omega_i +\alpha t)|\omega_s\rangle |\omega_i\rangle,
\end{align}
where $P(t)$ is a time dependent scaling factor determined by the shape of the pump pulse and $\alpha$ is the slope of the linear region of the resonance motion in time as shown in Fig. \ref{fig6}(b). The generated photons in this case are chirped. The shape of the optical pulse inside the cavity will match the input pulse shape as long as the amplitude variations are slow which we have already assumed by setting $\Omega\ll\gamma_t$.

\section{Discussion and Conclusion}
Complex quantum states such as entanglement in many dimensions or between different degrees of freedom or particles have been shown to be a potent tool for quantum technologies \cite{review2,review3,review4}. For instance, higher dimensional entangled states have been demonstrated to reduce the complexity of computation and increase density of encoding \cite{simplifying,highD}. These states, in principle, allow the computation of problems or simulation of systems that are cumbersome or otherwise intractable with current technologies. However, the realization of these schemes is a daunting task due to the inherent complexity of large quantum states and often necessitates use of complex optical circuits or multiplexing many sources of quantum light. 
\par Our analysis has shown that complex higher dimensional and hyper-entangled states can be constructed by parametric interactions inside a dynamically modulated cavity and is easily implementable on a chip. The properties of these complex states can be controlled within the photon generation process. We have further shown that the generated higher dimensional state has features of quantum random walks, a phenomenon useful in quantum simulation and information processing. We have calculated different instances of the modulation signal that lead to localization and delocalization of the walk. This system can also be used as a tunable quantum frequency shifter for applications in spectroscopy with squeezed light \cite{squeeze} when the wavefunction is localized and the frequency of the localization can be adjusted by the modulation signal's base frequency $\Omega$. This also makes the system appealing as a heralded single photon source with an adjustable frequency as it does not need quantum frequency conversion protocols to bridge bandwidth gaps between different optical systems when the frequency differences are small. In the case of electro-optic modulation, the achievable frequency difference is limited only by the electronics.

\par On a fundamental note, we have investigated emission of non-classical light in a dynamic environment and studied how the emission process is subsequently modified.  We envision that systems similar to the one presented in this article will prove to be of great interest in quantum simulation and information processing in the time-frequency domain.

\begin{acknowledgments}
This work was done with support from National Science Foundation Grant No. ECCS-1842691, ECCS-1351697 and EFMA-1641099.
\end{acknowledgments}

\appendix
\section{Tunneling rates of photons between adjacent bins} \label{AppA}
In this section we will show the equivalence of this system to a traditional system of many  coupled modes that is used for photonic random walks. This will directly give the coupling rates between adjacent frequency bins. 
\par Starting with a modified form of Eq. \eqref{eqs1} ignoring the external coupling and loss for simplicity and using $f(t)=G_m \cos (m\Omega t+\phi_m)$, we get
\begin{align}
\label{eqinput}
    \frac{da}{dt}=iG_m \cos(m\Omega t+\phi_m)a,
\end{align}
where the mode $a$ is in a frame rotating with its center frequency. Plugging a Floquet solution of the form $a=\sum_n a_n e^{-in\Omega t}$, Eq. \eqref{eqinput} becomes
\begin{align}
\label{eqsum}
     &\sum_n \Big\{ \frac{da_n}{dt}-in\Omega a_n \Big\} e^{-in\Omega t} \nonumber \\
     &=iG_m \cos(m\Omega t+\phi_m) \sum_n a_n e^{-in\Omega t} \nonumber \\
     &=i\frac{G_m}{2} \Big\{ e^{i\phi_m}\sum_n a_n e^{-i(n-m)\Omega t}+e^{-i\phi_m}\sum_n a_n e^{-i(n+m)\Omega t}\Big\}.
\end{align}
Redefining the indices of the sums on the right hand side of Eq. \eqref{eqsum}, we get
\begin{align}
     =i\frac{G_m}{2}\sum_n \Big\{ e^{i\phi_m} a_{n+m}+e^{-i\phi_m} a_{n-m} \Big\} e^{-in\Omega t}.
\end{align}
Comparing the terms proportional to the n-th harmonic of the modulation frequency, we get
\begin{align}
\label{eqA1}
     \frac{da_n}{dt}=in\Omega a_n+i\frac{G_m}{2} \Big\{ e^{i\phi_m} a_{n+m}+e^{-i\phi_m} a_{n-m} \Big\} .
\end{align}
Eq. \eqref{eqA1} represent a system of coupled modes with a coupling or tunneling strength $G_m/2$. Taking an analogy to the traditional coupled waveguide systems used for random walks, the n-th harmonic of the modulation frequency acts as an effective propagation constant mismatch while the modulation amplitude of the harmonic that connects adjacent frequency bins represents the strength of the mode overlap between adjacent waveguides.

\bibliography{apssamp}

\end{document}